# Metaverse Support Groups for LGBTQ+ Youth: An Observational Study on Safety, Self-Expression, and Early Intervention

Joe HASEI
*Department of Medical Information and Assistive Technology Development*
*Graduate School of Medicine*
*Okayama University*
*Okayama, Japan*
*py3g9rcw@s.okayama-u.ac.jp*
0000-0002-4028-0786
(Corresponding Author)

Yosuke MATSUMOTO
*Department of Neuropsychiatry*
*Graduate School of Medicine*
*Okayama University*
*Okayama, Japan*
*ymatsumo@okayama-u.ac.jp*
0000-0002-5538-0827

Hiroki KAWAI
*Department of Neuropsychiatry*
*Graduate School of Medicine*
*Okayama University*
*Okayama, Japan*
*pq8d2zxk@okayama-u.ac.jp*
0000-0002-4441-9896

Yuko OKAHISA
*Department of Neuropsychiatry*
*Graduate School of Medicine*
*Okayama University*
*Okayama, Japan*
*okahis-y@cc.okayama-u.ac.jp*
0000-0003-1125-0761

Manabu TAKAKI
*Department of Neuropsychiatry*
*Graduate School of Medicine*
*Okayama University*
*Okayama, Japan*
*manabuta@cc.okayama-u.ac.jp*
0000-0002-7371-2821

Toshifumi OZAKI
*Science of Functional Recovery and Reconstruction*
*Graduate School of Medicine*
*Okayama University*
*Okayama, Japan*
*tozaki@md.okayama-u.ac.jp*
0000-0003-1732-9307

*Abstract*—This study explored whether metaverse-based support groups could address social isolation and suicide risks among LGBTQ+ youths by providing enhanced anonymity, avatar-based self-expression, and improved accessibility. Over one year, 53 individuals aged 14–23 participated in regular online sessions facilitated via the "cluster" metaverse platform by a non-profit LGBTQ+ organization. Each 90-minute session included voice and text-based interactions within a specially designed single-floor virtual space featuring conversation areas and a designated "safe area" for emotional regulation. Post-session questionnaires (5-point Likert scales) captured demographics, avatar preferences, self-confidence, and perceived safety, self-expression, and accessibility; responses were analyzed with Pearson's chi-squared test and Mann–Whitney U tests ($\alpha=0.05$). Results indicated that 79.2% of participants selected avatars aligned with their gender identity, reporting high satisfaction (mean = 4.10/5) and minimal discomfort (mean = 1.79/5). Social confidence was significantly higher in the metaverse compared with real-world settings ($p<0.001$), particularly among those with lower real-world confidence, who exhibited an average gain of 2.08 points. Approximately half of all first-time participants were aged 16 years or younger, which suggested the platform's value for early intervention. Additionally, the metaverse environment was rated significantly higher in safety/privacy (3.94/5), self-expression (4.02/5), and accessibility (4.21/5) compared with the real-world baseline, and 73.6% reported they felt more accepted virtually. However, some participants who had high confidence offline experienced mild adaptation challenges (mean decrease of 0.58 points), which highlighted that metaverse-based support may be more effective as a complement to in-person services rather than a replacement. Overall, these findings demonstrate that metaverse-based support groups can reduce psychological barriers for LGBTQ+ youth by facilitating safe and affirming virtual environments. The metaverse may help alleviate emotional distress and prevent further severe outcomes, such as suicidal ideation by providing early intervention, especially for adolescents unable to access conventional in-person services. Further research should examine its integration with existing clinical, community, and educational resources to ensure comprehensive, long-term support.

I. INTRODUCTION

Although understanding of sexual minorities (LGBTQ+) in society has gradually progressed recently, individuals within these populations, particularly adolescents and young adults, continue to experience significant social isolation and mental health challenges, necessitating urgent intervention and support[1, 2]. Studies have documented that prolonged social isolation and alienation not only exacerbate depressive symptoms and anxiety, but also significantly increase the risk of suicidal ideation in severe cases [3–6]. Almost 42% and 29% of LGBTQ+ adolescents have seriously considered suicide and attempted suicide at least once within the past six months, respectively [7]. In Japan, approximately 48% and 14% of LGBTQ+ youth have considered and attempted suicide, respectively, which indicates a significant issue. These objective data clearly demonstrate that LGBTQ+ youth experience serious mental health risks [8]. Notably, previous studies have reported that when individuals have someone to consult or a supportive environment to share their concerns—thereby reducing feelings of isolation and loneliness—the incidence of suicidal ideation and self-harm behaviors can decrease, suggesting that the metaverse may also contribute to mitigating suicide risk [9–11].

Multiple social structural factors underlie this situation. Primarily, understanding regarding sexual diversity in school

**Cite (APA):** Hasei, J., Matsumoto, Y., Kawai, H., Okahisa, Y., Takaki, M., Ozaki, T. (2025). Metaverse Support Groups for LGBTQ+ Youth: An Observational Study on Safety, Self-Expression, and Early Intervention. *Journal of Metaverse.* 5(2), 156-167. Doi: 10.57019/jmv.1639701

educational settings and within families is lacking [12–16]. Many educational institutions lack comprehensive educational programs regarding LGBTQ+ issues; furthermore, in many cases, individuals have inadequate understanding or support within their families. Second, there are psychological and social risk s associated with coming out to those around them. Individuals are often unable to disclose their sexual orientation or gender identity, particularly during adolescence and young adulthood, due to fear of bullying and discrimination, which can lead to a tendency to internalize their struggles [17, 18]. LGBTQ+ students experience disproportionate victimization in schools. According to a national survey conducted in the United States, 86.3% of students reported experiencing harassment based on their sexual orientation, gender expression, or gender identity [19]. Third, geographical and social access barriers may exist due to the concentration of support organizations and counseling centers in urban areas [20, 21]. These interconnected factors can lead to further pronounced feelings of isolation and loneliness, particularly among young people in rural areas or those without understanding and support within their families.

Regarding these complex challenges, "metaverse" spaces utilizing virtual reality (VR) technology have been gaining attention recently as a new potential solution. In the metaverse, participants can interact online through avatars (their digital representations) and take advantage of an environment that offers greater anonymity compared with the real world [22]. Recent literature characterizes the metaverse as a highly immersive virtual space, enabling real-time social interactions through customizable avatars, thus overcoming traditional geographic and societal barriers. Recent literature provides a comprehensive technological framework for the metaverse, identifying critical domains such as artificial intelligence, extended reality, and blockchain [23]. These integrated technologies collectively enable immersive virtual experiences, secure and flexible identity expression through avatars, and inclusive social interactions beyond geographic and societal constraints, thus presenting significant potential for addressing psychological and social isolation among marginalized groups such as LGBTQ+ youth. Events and support activities within the metaverse targeting the LGBTQ+ community are being actively developed; recent empirical evidence further illustrates that social virtual reality environments provide unique advantages for support groups, particularly within LGBTQ+ communities, by enabling embodied interactions and emotional exchanges that significantly reduce isolation and enhance authentic self-expression [24]. However, support using the metaverse, specifically focusing on the unique challenges faced by adolescents and young adults, has not yet been sufficiently established. Support utilizing the metaverse offers three primary advantages to overcome challenges difficult to address with traditional face-to-face support. First, it significantly reduces geographical limitations as it can be accessed from anywhere with an Internet connection. This enables young people in rural areas and those with difficulty participating in ongoing support programs. Second, it can ensure a high degree of anonymity and psychological safety, allowing individuals hesitant or anxious about coming out to participate with peace of mind. Third, individuals can freely express their desired appearance, voice, and gender identity via their avatars without being restricted by their physical appearance, voice, or legal sex. Thus, the metaverse can create an environment that can reduce psychological distress related to gender dysphoria and foster self-affirmation [25, 26].

Japan's cultural familiarity with virtual avatars and digital self-representation, influenced by anime and manga traditions, may provide a particularly conducive environment for metaverse-based interventions. The widespread acceptance of avatar-based interaction in Japanese digital culture potentially reduces barriers to authentic self-expression in virtual spaces, making it an ideal context for exploring metaverse applications in mental health support.

This study focused on the characteristics of the metaverse and, to the best of our knowledge, aimed to conduct the world's first regular support group meetings in a metaverse space exclusively for LGBTQ+ youths and quantitatively examine its clinical effectiveness and practical challenges. Specifically, the analysis was conducted from three perspectives: the degree of freedom in gender expression through avatars and its psychological impact, changes in self-confidence during interpersonal interactions compared with the real world, and a comprehensive evaluation of the metaverse environment regarding safety, self-expression, and accessibility. These findings can provide concrete strategies for effectively utilizing new digital technologies in comprehensive support measures aimed at reducing serious isolation and suicide risks among young people.

## II. METHODOLOGY

### A. Participants

This observational study employed a cross-sectional design to quantitatively assess the experiences of LGBTQ+ youth in metaverse-based support groups. The term "observational" refers to the epidemiological definition of non-interventional research design. Data were collected through structured post-session questionnaires administered over a one-year period.

This observational study included participants in a support group for LGBTQ+ youth held in a metaverse space over a one-year period from January 2024 to January 2025. During this period, a total of 21 support group sessions were conducted, with each session lasting approximately 90 minutes. This study was conducted in collaboration with Niji-zu (where "Niji" means "rainbow" in Japanese), a Tokyo-based non-profit organization (NPO) established in August 2016 to support LGBTQ+ youth aged 10–23 years. Niji-zu now operates nationwide and focuses on creating connections that enable LGBTQ+ children and young people to safely navigate their adolescence. It strives to create inclusive spaces where diverse sexual orientations and gender identities and all forms of diversity are accepted as commonplace, and work toward alleviating anxiety and isolation among youth.

Participants were recruited through online announcements via Niji-zu and existing community networks. During the study period, 53 youths (aged 14–23 years, mean age: 18.2) participated, and questionnaire responses were collected after the support group meetings.

*B. Metaverse Platform and Space Design*

The metaverse space was specifically designed through the collaborative efforts of LGBTQ+ support professionals and metaverse designers. The virtual environment was developed using Unity as a world within "cluster," a metaverse platform provided by Cluster, Inc. (Shinagawa, Tokyo, Japan). This platform was selected for several key reasons: its accessibility across multiple devices (smartphones, tablets, PCs, VR headsets), robust privacy settings, and a user-friendly interface suitable for young participants. Niji-zu operates multiple support centers across the country and employs staff who are geographically dispersed. For each metaverse session, staff members were randomly selected from among these various centers, ensuring a diverse and distributed facilitation team. Meanwhile, participants were free to join the sessions from their homes or any location with internet access, irrespective of whether they had previously attended face-to-face support at any particular center. Both staff and participants used nicknames during the sessions, which further minimized the likelihood of recognizing one another even if they had met before.

The space design was created through close collaboration with Niji-zu. It was designed with careful consideration of both the safety and comfort for LGBTQ+ participants. A single-floor layout was implemented without blind spots, which enabled staff to maintain appropriate oversight while ensuring participant privacy. The space incorporated multiple conversation areas, with carefully calculated distances between zones to prevent audio overlap, which allowed for private group discussions (Figure I).

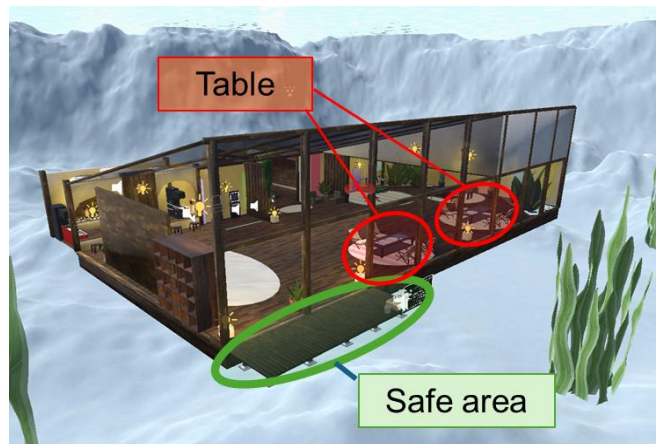

FIGURE I. DESIGN AND IMPLEMENTATION OF THE METAVERSE SPACE FOR LGBTQ+ YOUTH SELF-HELP GROUP.
*The space was designed as a single-floor structure with no blind spots, which allowed staff to monitor the entire area. Tables were positioned to allow for sub-group conversations, with distances optimized to minimize audio overlap from other groups.*

Additionally, a designated "Safe Area" was integrated, which featured calming elements, such as aquatic scenes with fish and jellyfish. This dedicated space was intentionally positioned outside the main interaction area and provided participants with an option to temporarily withdraw from social interaction when required. Its concept was specifically implemented to ensure that participants would have a legitimate and non-stigmatized space for emotional regulation and quiet reflection, which was an acknowledgement that social interactions, even in virtual spaces, could become overwhelming (Figure II).

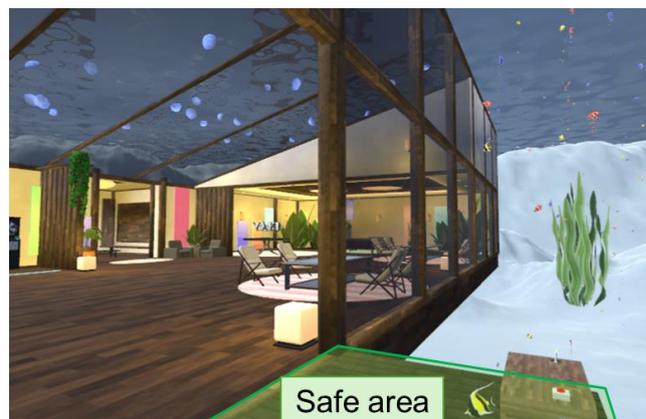

FIGURE II. DESIGN AND IMPLEMENTATION OF THE METAVERSE SPACE FOR LGBTQ+ YOUTH SELF-HELP GROUP
*A designated "safe area" was created outside the main room, which provided a retreat for participants who wished to take a break from the conversation. It featured calming visuals, such as fish and jellyfish.*

To ensure the support staffs' clear identification, team members used distinctive cat-style avatars, which made them easily recognizable to the participants. The space also incorporated interactive elements, such as casual games and activities, which included darts and board games, which served as natural conversation starters and helped facilitate comfortable social interactions. This design approach aimed to create an environment in which participants could engage at their own comfort level (Figure III).

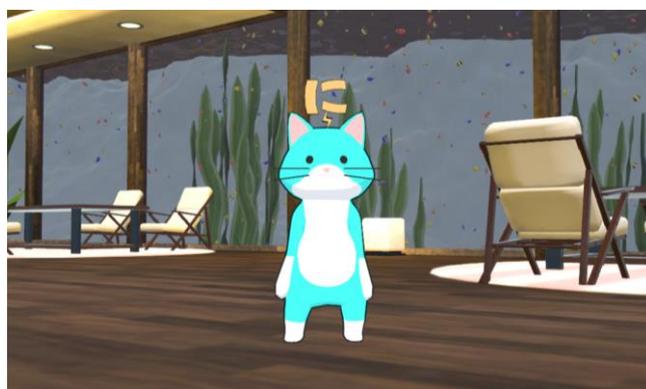

FIGURE III. DESIGN AND IMPLEMENTATION OF THE METAVERSE SPACE FOR LGBTQ+ YOUTH SELF-HELP GROUP.
*Staff members used distinctive cat avatars to be easily identifiable by the participants.*

The virtual environment underwent several refinements based on feedback from Niji-zu's experienced staff and preliminary user testing. Particular attention was paid to create an environment that would minimize any potential triggers of gender dysphoria and maximize opportunities for authentic self-expression. Niji-zu's extensive experience in creating safe spaces in which young people can freely express themselves, without fear of judgment or discrimination, has been instrumental in developing an effective virtual support environment. Their understanding of the unique challenges faced by LGBTQ+ youths, especially those who have not come out to their families or peers, proved invaluable in designing a space that could meaningfully address these sensitive needs and maintain participant safety and comfort (Figure IV).

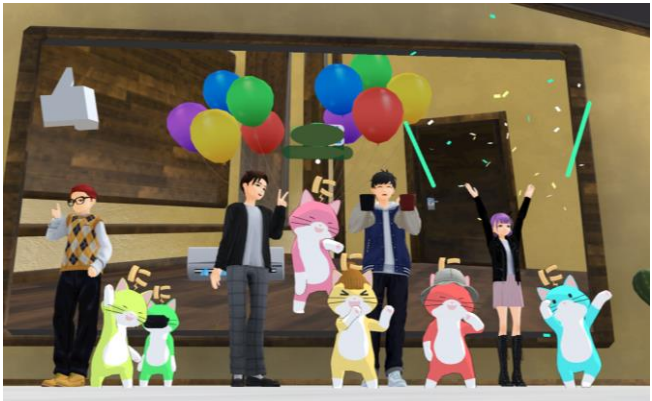

FIGURE IV. DESIGN AND IMPLEMENTATION OF THE METAVERSE SPACE FOR LGBTQ+ YOUTH SELF-HELP GROUP.
*The self-help group incorporated various interactive elements, such as games (darts, Othello), to facilitate communication and engagement. During the self-help group, participants enjoyed chatting about various topics and playing games.*

### C. Implementation Protocol

The researcher served strictly as a non-participant observer throughout the study period, without engaging in or influencing session dynamics, discussions, or participant interactions. All sessions were managed and facilitated exclusively by trained NPO staff members. Each support group session lasted approximately 90 minutes, divided into three 30-minute sessions, and included three to four Niji-zu staff members as facilitators. The sessions generally followed the steps outlined below.

#### 1) Identity Verification

To ensure safety and security, participants were asked to upload documents that verified their identity, such as student ID cards, to a secure cloud storage system. Subsequently, these documents were verified by Niji-zu staff. Participants were also required to pre-register the account information they would use in the session.

#### 2) Orientation

Once the participants' account information was confirmed to match their pre-registered information, they were welcomed into the metaverse. A brief explanation of the controls, safety guidelines, and rules of conduct was provided. The participants were informed that failure to comply with these rules would result in their removal.

#### 3) Themed Talk, Free Talk, or Activity

To facilitate icebreaking, participants introduced themselves along with their answers to simple questions posed by the staff (e.g., "What is your favorite winter food?" or "What do you want to do during summer vacation?"). Communication was enabled through voice and text chats. Interactive elements, such as darts and Othello, were also available in the virtual space, which provided opportunities for participants to enjoy games together and fostered communication.

#### 4) Q&A and Wrap-up

Before the end of each session, the staff and participants shared their impressions, and the staff solicited feedback regarding the participants' hopes for future sessions and suggestions for improvement.

### D. Data Collection Procedure

Data were collected through an online questionnaire administered after each metaverse session and obtained the following information:

- Basic Demographics: Age, gender, gender identity, and sex assigned at birth (AMAB/AFAB)
- Participation Status: First-time participation status, motivation for participation, and device used
- Self-Confidence Assessment: Level of confidence in social interactions in the real world and metaverse
- Avatar-Related Factors: Choice of gender expression (legal sex, gender identity, or other), level of satisfaction with self-expression, and any sense of incongruity
- Evaluation of the Metaverse Environment: Subjective ratings of safety, self-expression, and accessibility

Most evaluative items employed 5-point Likert scales, while demographic and preference questions used appropriate single or multiple-choice formats.

### E. Data Analysis

#### 1) Analysis of Participants' Demographics and Time Periods

The study period (January 25, 2024, to January 9, 2025) was divided into three equal periods based on the number of days (Period 1: January 25 to May 21, 2024; Period 2: May 21 to September 14, 2024; Period 3: September 14, 2024, to January 9, 2025). The number of participants and age distribution (≤16 years and >16 years) were compared across each period. A chi-squared test examined the differences in the number of participants across the periods. A binomial test was used to assess any increase in the number of participants in Period 3. Furthermore, differences in participant ratios between the adjacent periods (Period 1 vs. 2 and Period 2 vs. 3) were analyzed via Pearson's chi-squared test to compare proportions (significance level: 5%).

#### 2) Comparison of First-Time Participants and Other Participants

Participants were classified into two groups: first-time participants in metaverse support activities (those who answered, "I have never participated in any support group before (this is my first time)"), and others who had prior experience with support groups or interventions. Age distribution and motivation for participation were compared between the two groups. Regarding age, the difference in proportions was examined via a Pearson's chi-squared test, with 16 years as the threshold and a significance level of $p < 0.05$.

*3) Avatar Expression Choices and Psychological Indicators*

A chi-squared test was used to examine the distribution of avatar gender expression choices. Furthermore, a 2 x 2 chi-squared test was performed to compare the group that prioritized gender identity when they chose an avatar compared with the group that used their legal sex or other options. Bonferroni correction (significance level α = 0.025) was applied to adjust for multiple comparisons. Levels of satisfaction with self-expression and feelings of incongruity (both measured on a 5-point scale) were compared between the groups via the Kruskal-Wallis and Mann-Whitney U tests, as appropriate.

*4) Comparison of Self-Confidence in Real-World and Metaverse Interactions*

Participants were divided into two groups based on their self-confidence scores in real-world interactions (A scores): low-score (A scores of 1–2, n = 26) or high-score groups (A scores of 4–5, n = 12). Change in self-confidence scores (B-A), calculated as the difference between metaverse (B scores) and real-world interaction self-confidence scores (A scores), was compared between the groups via a Mann-Whitney U test. Statistical analyses were performed via Python, and $p < 0.05$ was considered statistically significant.

*5) Evaluation of Safety, Self-Expression, and Accessibility*

For the three indicators of safety/privacy, self-expression, and accessibility (all measured on a 5-point scale), a score of "3" was theoretically considered equivalent to the real world. One-sample t-tests were conducted to compare the metaverse scores against the theoretical values. Additionally, participants' choices between the real world and metaverse as the place where they felt "truly accepted for who they are" were tabulated. A chi-squared test was used to examine any differences.

*F. Ethical Considerations*

This study was approved by the Institutional Review Board of the implementing institution (Approval Number: 2503-030). Given that this study did not handle personally identifiable information, informed consent was carefully managed using an opt-out procedure, clearly informing participants of their right to withdraw at any time without repercussions. Additional privacy protection measures were specifically implemented for minor participants. All participants accessed sessions using anonymous user accounts and were identified exclusively by nicknames. Any sensitive personal information collected during identity verification was securely stored and accessible only by authorized NPO staff. Moreover, considering that minor participants might not have disclosed their gender identity or sexual orientation to guardians, the opt-out approach was particularly important to safeguard participants' privacy.

## III. RESULTS

*A. Participant Demographics and Age Distribution*

A metaverse-based social interaction session was conducted with 53 participants. Participants' ages ranged from 14–23 years (Mean = 18.2 years). The age distribution exhibited a bimodal pattern with peaks at ages 16–17 (n=20) and 19 years (n=11). Participant distribution by gender identity categories revealed that those with binary gender identities ("Male," "Female," "More likely Male," and "More likely Female") and non-binary identities ("Neither Male nor Female," "Both Male and Female," and "Varies depending on situation") had a mean age of 18.3 (n=30) and 18.5 years (n=17), respectively. Participants uncertain of their gender identity or preferred not to specify (n=6) were notably younger, with a mean age of 16.5 years and were exclusively in the 16–17 age range (Figure V).

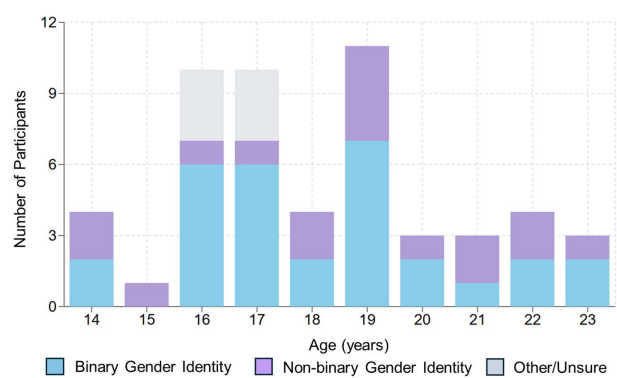

FIGURE V. PARTICIPANTS' DEMOGRAPHIC DISTRIBUTION BY AGE AND GENDER IDENTITY
*Age distribution (n=53) categorized by gender identity type. Blue, purple, and gray bars represent binary gender identities (Male, Female, More likely Male, More likely Female), non-binary identities (Neither Male nor Female, Both Male and Female, Varies depending on situation), and Other/Unsure responses, respectively. The distribution peaked at ages 16–17 (n=20) and 19 years (n=11).*

*B. Distribution of Gender Identity and Sex Assigned at Birth*

Among the 53 participants, the most common gender identity was "More likely Male" (n=13, 24.5%), with the majority being Assigned Female at Birth (AFAB) (n=10, 76.9%). Those who identified as "Female" constituted the second-largest group (n=9, 17.0%), with an almost equal distribution between Assigned Male at Birth (AMAB) (n=4) and AFAB (n=5). Non-binary gender identities represented a substantial portion (n=17, 32.1%), which included those who identified as "Neither Male nor Female" (n=7), "Both Male and Female" (n=7), and "Varies depending on situation" (n=3). In these non-binary groups, AFAB individuals were more prevalent. Overall, AFAB individuals were more represented across most gender identity categories. The total distribution revealed 35 AFAB (66.0%), nine AMAB (17.0%), and nine (17.0%) who preferred not to disclose their sex assigned at birth (Figure VI). Data revealed that while some respondents identified strongly with binary gender categories (Male/Female), a substantial portion chose more nuanced options, such as "More likely Male/Female" or non-binary identities, which suggested a complex spectrum of gender experiences.

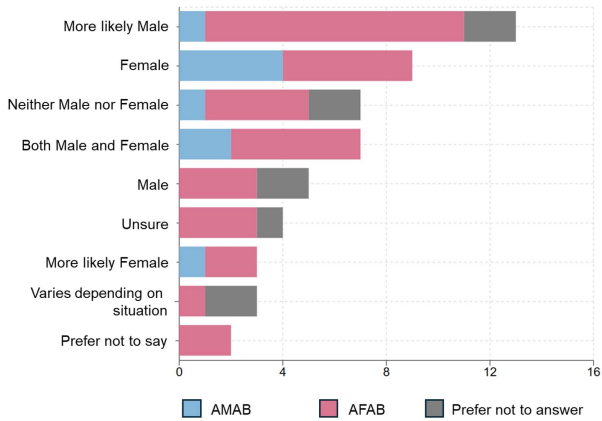

FIGURE VI. PARTICIPANTS' DEMOGRAPHIC DISTRIBUTION BY AGE AND GENDER IDENTITY
*Distribution of gender identities by sex assigned at birth. Horizontal bars represent the number of participants for each gender identity category, subdivided by AMAB (Assigned Male at Birth, blue), AFAB (Assigned Female at Birth, pink), and those who preferred not to answer (gray). "More likely Male" was the most common identity (n=13, 24.5%), with AFAB individuals representing the majority across most categories (n=35, 66.0%).*

*C. Participants' Demographics and Temporal Trends*

Of the 53 participants, 15 (five aged ≤16 and 10 aged >16 years), 14 (three aged ≤16 and 11 aged >16 years), and 24 (seven aged ≤16 and 17 aged >16 years) participated in Periods 1, 2, and 3, respectively. The number of participants in Period 3 (n = 24) was significantly higher than in Period 2 (n = 14) (P = 0.043), which accounted for 45.3% of all participants. Regarding age distribution, participants aged >16 years were predominant throughout all periods (66.7%–78.6%). Although the absolute number of participants aged ≤16 years was highest in Period 3 (n = 7), no significant change was observed in their proportion across the periods (Period 1:33.3%, Period 2:21.4%, Period 3:29.2%) (Figure VII).

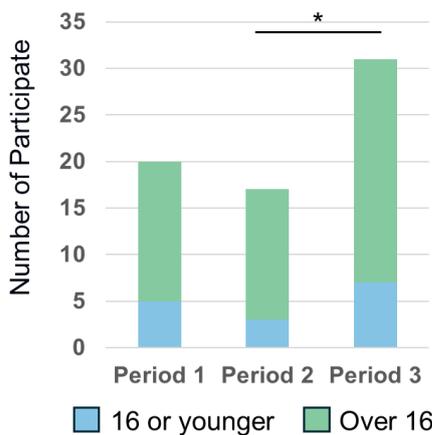

FIGURE VII. TEMPORAL CHANGES IN AGE DISTRIBUTION AND COMPARISON OF FIRST-TIME VS. OTHER PARTICIPANTS
*Stacked bar chart represents the number of participants in the three defined time periods (Period 1, 2, 3), categorized by age group (16 or younger in blue, over 16 in green). An asterisk (*) indicates a significant increase in the total number of participants from Period 2 to 3 (Pearson's Chi-squared test; p = 0.043).*

In this study, we classified participants into two groups based on their self-reported prior experience with Niji-zu's support: (1) "first-time," who had never previously attended any face-to-face or online support activities provided by Niji-zu, and (2) "more than two times," who had attended at least one such activity in the past. Thus, "more than two times" in the questionnaire referred to prior Niji-zu support sessions in general, not repeated attendance within this specific metaverse study. Analysis of the first-time participant group (n = 12) revealed an equal distribution between participants aged ≤16 (50.0%, n = 6) and >16 years (50.0%, n = 6). In contrast, the other participant group (n = 41) revealed a distribution skewed toward older age groups, with 22.0% (n = 9) aged ≤16 and 78.0% (n = 32) aged >16 years. This difference in, which used 16 years as the threshold, was statistically significant ($p = 0.048$), which confirmed a significantly higher proportion of younger participants in the first-time participant group (Figure VIII).

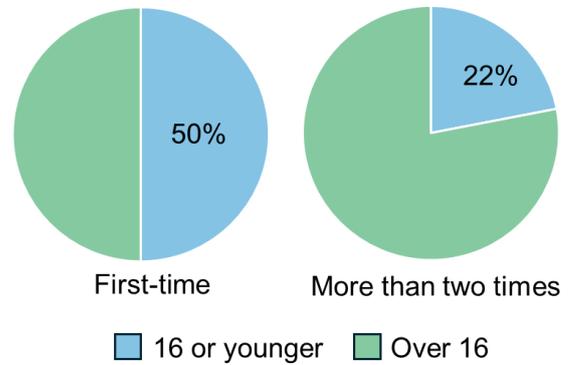

FIGURE VIII. TEMPORAL CHANGES IN AGE DISTRIBUTION AND COMPARISON OF FIRST-TIME VS. OTHER PARTICIPANTS
*Pie charts illustrate the proportion of participants aged 16 or younger (blue) versus over 16 (green) among first-time attendees (left) and those who participated more than twice (right). In the first-time group, the two age categories were evenly split (50–50), whereas the repeated-attendance group exhibited a distribution of 22% (16 or younger) and 78% (over 16). This difference was statistically significant (Z = 1.897, p = 0.048, two-tailed test), which indicated a higher proportion of younger participants among first-time attendees.*

*D. Avatar Gender Expression Preferences*

Analysis of avatar gender expression preferences revealed similar patterns between participants with gender dysphoria (n=18) and those who identified as non-binary (n = 17). Among those with gender dysphoria, 94.4% (17/18) chose to express their experienced gender identity through their avatars, while only 5.6% (1/18) opted to use their legal gender. Similarly, in the non-binary group, 88.2% (15/17) chose to express their experienced gender identity, while 11.8% (2/17) used their legal gender (Figure IX). This suggested that both groups exhibited a strong and comparable preference for expressing their experienced gender identity in the metaverse, regardless of whether they had experienced gender dysphoria or identified as non-binary. Among the 53 respondents, 42 (79.2%), six (11.3%), and five (9.4%) chose avatars that represented their experienced gender (gender identity), 6 (11.3%) were congruent with both their gender identity and legal gender, and represented their legal gender, respectively. The mean satisfaction score for self-expression was 4.10 (SE

= ±0.12), 4.33 (SE = ±0.33), and 4.20 (SE = ±0.37) for the group that chose avatars that represented their experienced gender, were congruent with both their gender identity and legal gender, and represented their legal gender, respectively.

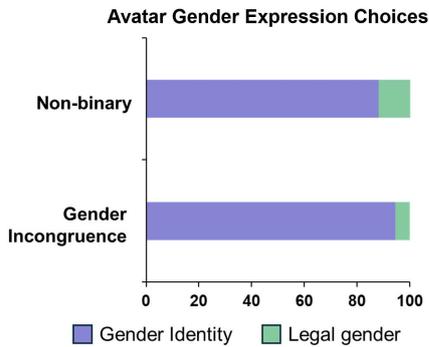

FIGURE IX. AVATAR GENDER EXPRESSION PREFERENCES AND ASSOCIATED PSYCHOLOGICAL METRICS.
*Distribution of avatar gender expression choices among non-binary individuals (n=17) and those with gender incongruence (n=18). Purple and green bars represent participants who chose avatars reflecting their gender identity (non-binary: 88.2%, 15/17; gender incongruence: 94.4%, 17/18) and legal gender (non-binary: 11.8%, 2/17; gender incongruence: 5.6%, 1/18), respectively.*

Regarding discomfort scores related to avatar gender expression, the mean score was 1.79 (SE = ±0.15), 1.50 (SE = ±0.34), and 1.40 (SE = ±0.25) for the Gender Identity Avatar, Congruent Gender Avatar, and Legal Gender Avatar groups, respectively. All three groups exhibited similarly low levels of discomfort, with no statistically significant differences (Figure X).

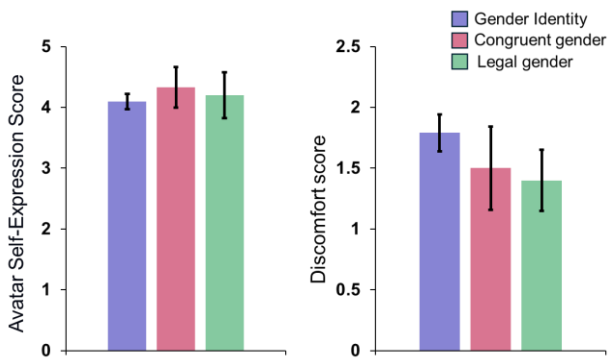

FIGURE X. AVATAR GENDER EXPRESSION PREFERENCES AND ASSOCIATED PSYCHOLOGICAL METRICS.
*Avatar-related psychological metrics by gender choice groups. Left: Self-expression satisfaction scores (5-point scale, higher = more satisfied). Right: Discomfort scores (5-point scale, higher = more discomfort). Groups shown: gender identity avatars (n=42, purple), congruent gender/legal gender avatars (n=6, pink), and legal gender avatars (n=5, green). Error bars indicate standard error. All groups exhibited high satisfaction (means 4.10–4.33) and low discomfort (means 1.40–1.78), with no statistically significant differences.*

### E. Comparison of Confidence Scores: Real World vs. Metaverse

Among all 53 participants, the mean self-confidence score for interacting with others in the real world and metaverse was 2.58 (SE = 0.17) and 3.60 (SE = 0.16), respectively. Overall, self-confidence in the metaverse was, on average, 1.02 points higher than in the real world, a statistically significant difference (Mann-Whitney U test; $p=6.50\times10^{-5}$) (Figure XI).

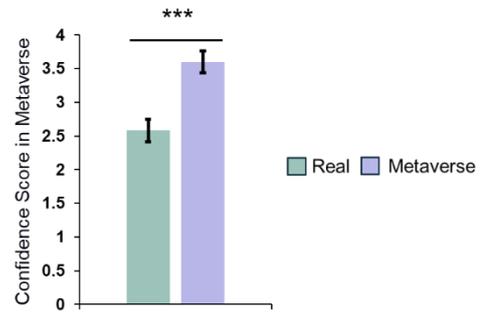

FIGURE XI. CONFIDENCE LEVELS IN REAL-WORLD VERSUS METAVERSE INTERACTIONS
*Comparison of mean confidence scores in real-world (green) and metaverse (purple) interactions. Error bars represent standard error; \*\*\*p < 0.001 by Mann-Whitney U test.*

Analysis of self-confidence scores in the real world and change in self-confidence within the metaverse revealed that most participants had a self-confidence score of 3 or lower in the real world (77.3%). Conversely, approximately 83.0% had a self-confidence score of 3 or higher in the metaverse. These results suggested that most participants interacted with others in the metaverse with a moderate or higher level of self-confidence (Figure XII).

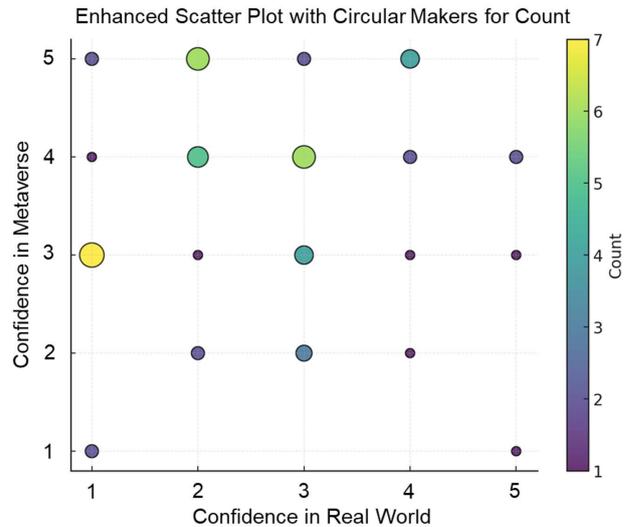

FIGURE XII. CONFIDENCE LEVELS IN REAL-WORLD VERSUS METAVERSE INTERACTIONS
*Correlation between real-world and metaverse confidence scores. Circle size and color indicate response frequency (yellow = higher, purple = lower).*

Interestingly, different patterns of change in self-confidence in the metaverse were observed between the low and high self-confidence groups in the real world. The group with low self-confidence in the real world (scores of 1-2) exhibited a mean increase of 2.08 points (SD = 1.13). Conversely, the group with the highest self-confidence (score of 5, n = 12) exhibited a mean decrease of 0.58 points (SD = 1.56). This difference was statistically significant (Mann-Whitney U test; $p=1.80\times10^{-5}$). These findings suggested that the metaverse may have a particularly positive effect on participants with low self-confidence in the real world (Figure XIII).

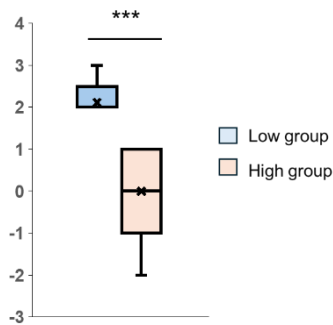

FIGURE XIII. CONFIDENCE LEVELS IN REAL-WORLD VERSUS METAVERSE INTERACTIONS
*Changes in confidence scores between real-world and metaverse environments. Box plots represent low (1–2 points, blue) and high (5 points, pink) real-world confidence groups. Low-confidence group demonstrated a mean increase of 2.08±1.13, while high-confidence group demonstrated a mean decrease of 0.58±1.56; ***p < 0.001 by Mann-Whitney U test.*

### F. Evaluation of Safety/Privacy, Self-Expression, and Accessibility in the Metaverse

The mean score for safety/privacy in the metaverse was 3.94 (SE = 0.14), which was statistically significantly higher than the theoretical score of 3 in the real world (t = 6.81, $p = 9.80 \times 10^{-9}$). Illustrative of this sentiment, one participant remarked, "*It was a very enjoyable event! I was initially a bit nervous about gathering in VR for the first time, but the environment felt very gentle and relaxing.*" Similarly, the mean self-expression score in the metaverse was 4.02 (SE = 0.14), which significantly exceeded the real-world score of 3 (t = 7.35, $p = 1.38 \times 10^{-9}$). One participant noted positively, "*It was really fun! I could relax and participate wearing exactly what I wanted. Thank you!*" Another participant highlighted deeper implications of this virtual freedom, stating, "*I felt more accepted as my true self in the virtual space, perhaps because it allowed me to mask gender differences arising from uncontrollable aspects like my voice pitch and physical build.*" The mean score for accessibility was even higher, at 4.21 (SE = 0.12) (t = 9.93, $p = 1.30 \times 10^{-13}$) (Figure XIV).

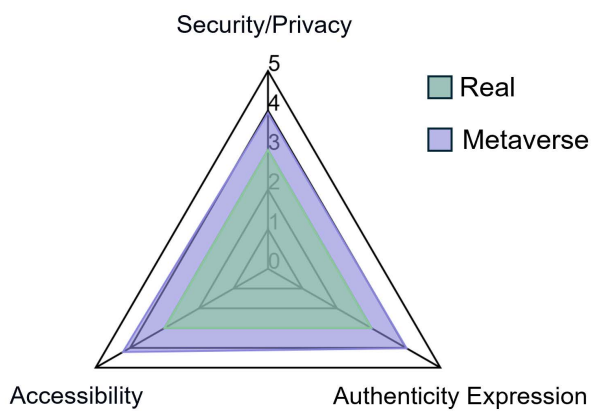

FIGURE XIV. EVALUATION METRICS COMPARING METAVERSE AND REAL-WORLD ENVIRONMENTS.
*Radar plot comparing the metaverse (purple) and real-world (green) environments across three dimensions: Security/Privacy, Authenticity Expression, and Accessibility. Scores range from 0–5, with real-world baseline set at 3. Metaverse scores were significantly higher for all metrics (p < 0.001, one-sample t-test).*

Furthermore, when asked to choose between the metaverse and real world as the place where they felt "truly accepted for who they are," 73.6% and 26.4% selected the metaverse and real world, respectively (Pearson's Chi-squared test; p = 5.95 × $10^{-4}$). This finding indicated that the metaverse may have high utility for LGBTQ+ youth regarding self-expression and acceptance (Figure XV).

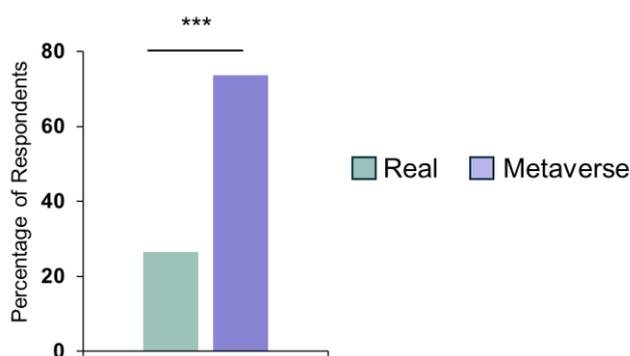

FIGURE XV. EVALUATION METRICS COMPARING METAVERSE AND REAL-WORLD ENVIRONMENTS
*Percentage of participants selecting the metaverse (73.6%, purple) versus real-world (26.4%, green) as the environment where they feel most accepted for their true selves; ***p < 0.001, Pearson's Chi-squared test.*

## IV. DISCUSSION

### A. Self-Expression and Psychological Safety in the Metaverse

This study was significant as it provided an initial examination of the psychological and social effects of support groups for LGBTQ+ youth utilizing a metaverse environment. Notably, most participants were able to express their authentic selves via avatars without being concerned of their physical appearance in the real world or legal gender. High levels of satisfaction and low levels of incongruity suggested that the metaverse offered a high degree of psychological safety to LGBTQ+ youths (Fig. 4B). Previous research indicated that young LGBTQ+ individuals often faced discrimination and prejudice in offline environments, such as schools and homes, which could lead to isolation before they fully established their self-esteem [27, 28]. In contrast, the anonymity and flexibility afforded by customizable avatars in the metaverse made it a space where individuals could positively express their gender identity and sexuality [26, 29]. Our results further showed that participants with gender dysphoria (94.4%) and non-binary individuals (88.2%) overwhelmingly selected avatars reflecting their experienced gender identity (Fig. 4A), suggesting that both groups shared a strong preference for authentic self-expression regardless of clinically recognized dysphoria.

One plausible explanation for these findings is that the metaverse setting reduces the external pressures associated with physical appearance or legal gender, thus enabling more authentic and affirming expressions of self.

In addition, Fig. 6A demonstrates that the metaverse may offer a safer and more conducive environment for self-expression among LGBTQ+ youth. The particularly high accessibility score highlights the metaverse's effectiveness in reducing geographical and psychological barriers. Furthermore, when asked to choose between the metaverse and the real world as the place where they felt "truly accepted for who they are," 73.6% and 26.4% selected the metaverse and the real world, respectively (Pearson's Chi-squared test; $p = 5.95 \times 10^{-4}$). This finding indicated that the metaverse may have high utility for LGBTQ+ youth regarding self-expression and acceptance (Fig. 6B).

By mitigating concerns about discrimination or judgment, the metaverse environment can serve as a psychologically safe venue where young LGBTQ+ individuals can experiment with and validate their identities.

### B. Effects and Limitations of Online Communication

The significant increase in self-confidence among participants who lacked confidence in real-world interactions, following their interactions in the metaverse, highlighted the advantages of online communication. Interactions with others in the real world often among LGBTQ+ youths during adolescence and young adulthood involve a significant psychological burden, which forces them to be constantly aware of others' gaze and criticism regarding their appearance, voice, and behavior [30, 31]. In contrast, our participants experienced a process of gradually building relationships through games, activities, and text chat. The metaverse, which offers a high degree of freedom from preconceptions based on appearance and voice, likely functioned as a useful space for fostering confidence in social interactions, which was unattainable through face-to-face interactions. Conversely, some participants who already possessed high self-confidence in the real world exhibited a slight sense of discomfort or a tendency to experience decreased self-confidence in the metaverse environment. This may be attributed to the limitations in online non-verbal communication, as well as initial unfamiliarity with avatar manipulation and voice settings. Therefore, metaverse-based support groups should not be considered a complete replacement for real-world support; rather, it should be a complementary approach, particularly for individuals with difficulty accessing face-to-face support or those who find real-world communication challenging.

### C. Potential of the Metaverse as an Early Intervention Tool

The finding that approximately half of the first-time participants were aged 16 years or younger suggested that metaverse-based support held significant potential for early intervention. Our findings align with previous studies demonstrating the positive impact of online and virtual support on LGBTQ+ youth mental health. For instance, Li et al. reported similar emotional relief and reduced isolation among LGBTQ+ individuals engaging in social VR environments [24]. Additionally, Berger et al. demonstrated that social media use for peer connection, identity management, and social support is associated with improved mental health and well-being in LGBTQ+ youths [32]. However, our study uniquely highlights the importance of immersive, avatar-based interactions provided by the metaverse, which may further enhance emotional engagement and community belonging compared to traditional text-based social media platforms. This emotional benefit was captured by one participant who stated: "*I enjoy it because I feel emotionally relieved during the event.*" The process of establishing one's sexual orientation and gender identity during adolescence can lead to severe isolation and suicidal ideation, particularly when lacking understanding from those around them [33, 34]. Therefore, appropriate support and interventions are crucial. However, traditional face-to-face support has significant barriers to participation, especially for younger individuals. The issue of coming out to their parents is the most notable. Fear of rejection creates a substantial psychological burden for many youths considering coming out [35, 36], and attending a support group meeting in person without informing their parents is often impossible. Therefore, the metaverse, which can be safely accessed from home, can be a powerful tool for supporting young people struggling with isolation and loneliness. Previous research revealed that participation in LGBTQ+ communities enhanced resilience and improved future outlook [37]. Furthermore, the use of social media contributed to improved mental health and well-being [32]. However, while social media primarily relies on text-based communication, such as chat and posts, the metaverse offers a stronger sense of presence and shared experience by allowing users to participate in the "same place" and "same events" together. This can contribute to a greater reduction in feelings of isolation and foster a stronger sense of community. Moreover, the metaverse can provide broader accessibility with its high degree of anonymity, diverse options for self-expression through avatars, voice modulation via voice changers, and availability of non-verbal communication tools, similar to social media, while further expanding the scope of anonymity and self-expression. Hence, the metaverse offers a promising avenue for LGBTQ+ youths to interact safely, making it a valuable tool for early intervention. While this study was conducted in Japan's relatively stable social context, it is important to acknowledge that global trends toward LGBTQ+ acceptance are not uniform. In regions experiencing setbacks in diversity and equity protections, metaverse-based support may become even more critical as a safe space for youth who face increasing real-world discrimination. This underscores the importance of developing robust virtual support systems that can provide psychological safety regardless of local political climates.

### D. Limitations and Future Directions

Due to the observational nature of this study, the extent to which metaverse-based support contributed to long-term mental health improvement and suicide risk reduction was not fully evaluated. Furthermore, the limited sample size necessitates further research to conduct longitudinal studies with larger and more diverse populations and rigorous methodologies with control groups to further validate the effectiveness and sustainability of metaverse-based support.

While our study utilized "cluster," a metaverse platform with functionality and accessibility typical of many contemporary platforms, exploring additional metaverse platforms in future studies could further confirm the generalizability and robustness of these results. Additionally, given our relatively small and self-selected sample, caution is advised in generalizing these findings broadly. Future research should therefore include larger, more diverse participant samples to strengthen generalizability and applicability.

Ensuring safety within the metaverse is also a critical issue. Specific concerns include preventing harassment and privacy violations as well as protecting minors' rights. Since many participants have not come out to their families regarding their sexual orientation or gender identity, the traditional informed consent process requiring parental consent presents inherent limitations. Therefore, a new framework that maximizes participants' autonomy while ensuring their safety is required. This could be achieved through collaboration among researchers, support organizations, and platform providers to develop effective solutions.

Furthermore, to maximize the benefits of metaverse-based support, o defining clear methods for collaboration with existing medical and welfare resources, schools, and local communities, is critical. Online support alone may not be sufficient in cases involving severe mental health issues or challenging family environments. However, leveraging the metaverse's flexibility and combining it with a multidisciplinary approach involving professionals can potentially provide appropriate support to young people with difficulty accessing traditional support modalities.

These findings suggest that the metaverse represents a powerful option for LGBTQ+ youths to gain psychological safety and freedom of self-expression, and potentially offers significant benefits, especially for younger individuals and those who lack confidence in real-world settings. Future implementation research can provide concrete strategies for building comprehensive support systems for young people from diverse backgrounds and for mitigating isolation and discrimination in society.

Lastly, while overall satisfaction was high, participants also identified important technical barriers that must be addressed for broader implementation. One participant described: "*My laptop had many school-related software installations, which made it very slow and challenging to participate.*" while another noted, "*I joined via smartphone, but the app felt heavy. It would be good if participation were easier for those without optimal IT environments.*" These technical limitations highlight the need for future improvements in platform optimization and device compatibility to ensure truly inclusive access to metaverse-based support interventions.

These findings offer several considerations for practice and policy. The demonstrated benefits of metaverse-based support suggest potential value as a complement to existing services for LGBTQ+ youth, particularly those facing access barriers to traditional resources. The technical challenges identified by participants underscore the importance of addressing digital equity and platform usability in virtual intervention design. These results may inform future development of virtual support services and consideration of technological infrastructure needs.

## V. CONCLUSION

This study, a world-first trial, implemented exclusive support group meetings in a metaverse space for LGBTQ+ youth and examined their effectiveness and challenges. Results revealed a high degree of freedom in self-expression through avatars, and participants who experienced difficulties in real-world interactions exhibited significantly improved self-confidence in virtual space. Furthermore, significantly higher evaluations of safety, self-expression, and accessibility in the metaverse compared with their theoretical real-world equivalents suggested that the metaverse held great potential as a new platform for LGBTQ+ youths to alleviate feelings of isolation, overcome geographical and psychological barriers, and express themselves with greater confidence.

However, as young people actively use online spaces, challenges persist regarding operational and security management, privacy protection, and countermeasures against online harassment. Moreover, since some cases involving severe mental health or family issues that cannot be adequately addressed solely through the metaverse, strengthening collaboration with existing medical and welfare institutions and professionals by combining multiple forms of support is necessary.

Although this study employed an observational framework, long-term verification with larger and more diverse populations is required. However, this is the first study to empirically demonstrate the feasibility and effectiveness of metaverse-based support, specifically for LGBTQ+ youth. Therefore, it carries significant implications. Building upon these findings, can enable the establishment of a more comprehensive support systems that effectively combine metaverse-based and real-world (face-to-face) support. Furthermore, advancements in digital technology and strengthened collaborative frameworks, including multidisciplinary approaches, can pave the way for concrete strategies to mitigate isolation and discrimination in society and improve the mental health and well-being of LGBTQ+ youths.

### ACKNOWLEDGEMENT

We express our deepest gratitude to all the LGBTQ+ youth who participated in this study and to NPO Niji-zu for their invaluable collaboration in organizing and facilitating the support group sessions. We also thank Cluster, Inc. for their technical cooperation in providing the "cluster" metaverse platform used in this study; their technology was instrumental in creating an accessible and engaging virtual environment for our participants.

### FUNDING

This research was supported by The Mitsubishi Foundation, Japan (Grant No. 202430037); Hashimoto Foundation Inc., Japan (FY2023); Children and Families Agency, Government of Japan (Grant No. 165).

### AUTHORS` CONTRIBUTIONS

Conceptualization, J.H., Y.M. and M.T.; methodology, J.H. and Y.M.; investigation J.H. and Y.M.; resources, J.H.,

H.K. and Y.O.; writing—original draft preparation, J.H.; writing—review and editing, J.H., Y.M., M.T., and T.O.; visualization, J.H. All authors have participated in drafting the manuscript. All authors read and approved the final version of the manuscript.

CONFLICT OF INTEREST

The authors certify that there is no conflict of interest with any financial organization regarding the material discussed in the manuscript.

DATA AVAILABILITY

The data supporting the findings of this study are available upon request from the authors.

ETHICAL STATEMENT

This article followed the principles of scientific research and publication ethics.

DECLARATION OF AI USAGE

No generative AI tools were used for content creation in this manuscript (e.g., drafting, rewriting, or generating ideas).

REFERENCES

[1] Garcia, J., Vargas, N., Clark, J. L., Magaña Álvarez, M., Nelons, D. A., & Parker, R. G. (2020). Social isolation and connectedness as determinants of well-being: Global evidence mapping focused on LGBTQ youth. *Global Public Health*, *15*(4), 497–519.

[2] Escobar-Viera, C. G., Choukas-Bradley, S., Sidani, J., Maheux, A. J., Roberts, S. R., & Rollman, B. L. (2022). Examining social media experiences and attitudes toward technology-based interventions for reducing social isolation among LGBTQ youth living in rural United States: An online qualitative study. *Frontiers in Digital Health*, *4*, 900695.

[3] Allred, A., & Allred, K. (2024). Saving lives: Suicide prevention in LGBTQ youth. *Journal of Psychosocial Nursing and Mental Health Services*, *62*(4), 6–8.

[4] Duncan, D. T., & Hatzenbuehler, M. L. (2014). Lesbian, gay, bisexual, and transgender hate crimes and suicidality among a population-based sample of sexual-minority adolescents in Boston. *American Journal of Public Health*, *104*(2), 272–278.

[5] Liu, R. T., & Mustanski, B. (2012). Suicidal ideation and self-harm in lesbian, gay, bisexual, and transgender youth. *American Journal of Preventive Medicine*, *42*(3), 221–228.

[6] DelFerro, J., Whelihan, J., Min, J., Powell, M., DiFiore, G., Gzesh, A., Jelinek, S., Schwartz, K. T. G., Davis, M., Jones, J. D., Fiks, A. G., Jenssen, B. P., & Wood, S. (2024). The role of family support in moderating mental health outcomes for LGBTQ+ youth in primary care. *JAMA Pediatrics*, *178*(9), 914–922.

[7] Hatchel, T., Ingram, K. M., Mintz, S., Hartley, C., Valido, A., Espelage, D. L., & Wyman, P. (2019). Predictors of suicidal ideation and attempts among LGBTQ adolescents: The roles of help-seeking beliefs, peer victimization, depressive symptoms, and drug use. *Journal of Child and Family Studies*, *28*(9), 2443–2455.

[8] ReBit. (2022). LGBTQ Children and Youth Survey 2022. *Certified NPO ReBit, Survey Report*, [Online]. Available: https://prtimes.jp/main/html/rd/p/000000031.000047512.html

[9] Green, A. E., Price-Feeney, M., & Dorison, S. H. (2021). Association of sexual orientation acceptance with reduced suicide attempts among lesbian, gay, bisexual, transgender, queer, and questioning youth. *LGBT Health*, *8*(1), 26–31.

[10] Gorse, M. (2022). Risk and protective factors to LGBTQ+ youth suicide: A review of the literature. *Child & Adolescent Social Work Journal: C & A*, *39*(1), 17–28.

[11] Eisenberg, M. E., Wood, B. A., Erickson, D. J., Gower, A. L., Kessel Schneider, S., & Corliss, H. L. (2021). Associations between LGBTQ+-supportive school and community resources and suicide attempts among adolescents in Massachusetts. *The American Journal of Orthopsychiatry*, *91*(6), 800–811.

[12] Fantus, S., & Newman, P. A. (2021). Promoting a positive school climate for sexual and gender minority youth through a systems approach: A theory-informed qualitative study. *The American Journal of Orthopsychiatry*, *91*(1), 9–19.

[13] Harris, R., Wilson-Daily, A. E., & Fuller, G. (2021). Exploring the secondary school experience of LGBT+ youth: an examination of school culture and school climate as understood by teachers and experienced by LGBT+ students. *Intercultural Education*, *32*(4), 368–385.

[14] Morgan, A., Cunningham, E., Dyrud, J., Elliott, L., Ige, L., Knowles, G., Konieczka, L., Mascolo, A., Sabra, I., Sabra, S., Singh, E., Rimes, K. A., & Woodhead, C. (2024). Intersectionality informed and narrative-shifting whole school approaches for LGBTQ+ secondary school student mental health: A UK qualitative study. *PloS One*, *19*(7), e0306864.

[15] Reczek, R., & Smith, E. B. (2021). How LGBTQ adults maintain ties with rejecting parents: Theorizing "conflict work" as family work. *Journal of Marriage and the Family*, *83*(4), 1134–1153.

[16] van Bergen, D. D., Wilson, B. D. M., Russell, S. T., Gordon, A. G., & Rothblum, E. D. (2021). Parental responses to coming out by lesbian, gay, bisexual, queer, pansexual, or two-spirited people across three age cohorts. *Journal of Marriage and the Family*, *83*(4), 1116–1133.

[17] Ojeda, M., Elipe, P., & Del Rey, R. (2024). LGBTQ+ bullying and cyberbullying: Beyond sexual orientation and gender identity. *Victims & Offenders*, *19*(3), 491–512.

[18] Bower-Brown, S., Zadeh, S., & Jadva, V. (2023). Binary-trans, non-binary and gender-questioning adolescents' experiences in UK schools. *Journal of LGBT Youth*, *20*(1), 74–92.

[19] Kosciw, J. G., Clark, C. M., Truong, N. L., & Zongrone, A. D. (2020). *The 2019 National School Climate Survey: The Experiences of Lesbian, Gay, Bisexual, Transgender, and Queer Youth in Our Nation's Schools. A Report from GLSEN*. https://eric.ed.gov/?id=ED608534.

[20] Cronin, T. J., Pepping, C. A., Halford, W. K., & Lyons, A. (2021). Mental health help-seeking and barriers to service access among lesbian, gay, and bisexual Australians. *Australian Psychologist*, *56*(1), 46–60.

[21] Wike, T. L., Bouchard, L. M., Kemmerer, A., & Yabar, M. P. (2022). Victimization and resilience: Experiences of rural LGBTQ+ youth across multiple contexts. *Journal of Interpersonal Violence*, *37*(19–20), NP18988–NP19015.

[22] Kiaer, J. (2024). *Conversing in the metaverse*. Bloomsbury Publishing Plc.

[23] Ghazinoory, S., Parvin, F., Saghafi, F., Afshari-Mofrad, M., Ghazavi, N., & Fatemi, M. (2025). Metaverse technology tree: a holistic view. *Frontiers in Artificial Intelligence*, *8*, 1545144.

[24] Li, L., Freeman, G., Schulenberg, K., & Acena, D. (2023). "we cried on each other's shoulders": How LGBTQ+ individuals experience social support in social virtual reality. *Proceedings of the 2023 CHI Conference on Human Factors in Computing Systems*, 1–16.

[25] Kundu, A., Barbareschi, G., Kawaguchi, M., Yano, Y., Ohashi, M., Kitaoka, K., Seike, A., & Minamizawa, K. (2024). "I wanted to create my ideal self": Exploring avatar perception of LGBTQ+ users for therapy in Virtual Reality. In *arXiv [cs.HC]*. arXiv. http://arxiv.org/abs/2409.00383.

[26] Povinelli, K., & Zhao, Y. (2024). Springboard, roadblock or "crutch"?: How transgender users leverage voice changers for gender presentation in social virtual reality. *ArXiv [Cs.HC]*. https://doi.org/10.48550/ARXIV.2402.08217.

[27] Hailey, J., Burton, W., & Arscott, J. (2020). We are family: Chosen and created families as a protective factor against racialized trauma and anti-LGBTQ oppression among African American sexual and gender minority youth. *Journal of GLBT Family Studies*, *16*(2), 176–191.

[28] Wilson, C., & Cariola, L. A. (2020). LGBTQI+ youth and mental health: A systematic review of qualitative research. *Adolescent Research Review*, *5*(2), 187–211.

[29] Morgan, H., O'Donovan, A., Almeida, R., Lin, A., & Perry, Y. (2020). The role of the avatar in gaming for trans and gender diverse young people. *International Journal of Environmental Research and Public Health*, *17*(22), 8617.

[30] Taylor, K., Coulombe, S., Coleman, T. A., Cameron, R., Davis, C., Wilson, C. L., Woodford, M. R., & Travers, R. (2022). Social support, discrimination, and Self-Esteem in LGBTQ + high school and Post-Secondary students. *Journal of LGBT Youth*, *19*(3), 350–374.

[31] Russell, S. T., Bishop, M. D., Saba, V. C., James, I., & Ioverno, S. (2021). Promoting school safety for LGBTQ and all students. *Policy Insights from the Behavioral and Brain Sciences*, *8*(2), 160–166.

[32] Weinhardt, L. S., Wesp, L. M., Xie, H., Murray, J. J., Martín, J., DeGeorge, S., Weinhardt, C. B., Hawkins, M., & Stevens, P. (2021). Pride Camp: Pilot study of an intervention to develop resilience and self-esteem among LGBTQ youth. *International Journal for Equity in Health*, *20*(1), 150.

[33] Bojarski, E., & Qayyum, Z. (2018). Psychodynamics of suicide in lesbian, gay, bisexual, or transgender youth. *Journal of Infant, Child, and Adolescent Psychotherapy: JICAP*, 1–9.

[34] Chang, C. J., Kellerman, J., Feinstein, B. A., Selby, E. A., & Goldbach, J. T. (2022). Greater minority stress is associated with lower intentions to disclose suicidal thoughts among LGBTQ + youth. *Archives of Suicide Research: Official Journal of the International Academy for Suicide Research*, *26*(2), 626–640.

[35] Sahoo, S., Venkatesan, V., & Chakravarty, R. (2023). 'Coming out'/self-disclosure in LGBTQ+ adolescents and youth: International and Indian scenario - A narrative review of published studies in the last decade (2012-2022). *Indian Journal of Psychiatry*, *65*(10), 1012–1024.

[36] Weinhardt, L. S., Wesp, L. M., Xie, H., Murray, J. J., Martín, J., DeGeorge, S., Weinhardt, C. B., Hawkins, M., & Stevens, P. (2021). Pride Camp: Pilot study of an intervention to develop resilience and self-esteem among LGBTQ youth. *International Journal for Equity in Health*, *20*(1), 150.

[37] Berger, M. N., Taba, M., Marino, J. L., Lim, M. S. C., & Skinner, S. R. (2022). Social media use and health and well-being of lesbian, gay, bisexual, transgender, and queer youth: Systematic review. *Journal of Medical Internet Research*, *24*(9), e38449.